\documentclass[twocolumn,aps,prb,a4paper,reprint,superscriptaddress,longbibliography]{revtex4-2}

\usepackage{graphicx}
\usepackage{multirow}
\usepackage[dvipsnames]{xcolor}
\usepackage{amsmath,amsfonts,amssymb}
\usepackage{xr-hyper}
\usepackage{physics}
\usepackage[vietnamese, english]{babel}
\usepackage{hyperref}
\usepackage{dsfont}
\usepackage[utf8]{inputenc}
\usepackage{soul}
\hypersetup{
     colorlinks   = true,
     allcolors    = blue,
}
\usepackage{array}

\begin{document}

%\title{N-layers rhombohedral graphene: An extensive Hartree-Fock study}
\title{Nematic and partially polarized phases in rhombohedral graphene with varying number of layers: An extensive Hartree-Fock Study} % Alex suggestion
%\title{Nematic, partial abnd polarized phases in rhombohedral $N$-layer graphene: An extensive Hartree-Fock Study} % Alex suggestion

\author{Guillermo Parra-Mart\'inez }
\thanks{These authors contributed equally.}
\affiliation{Fundaci\'{o}n IMDEA Nanociencia, C/ Faraday 9, Campus Cantoblanco, 28409 Madrid, Spain}
\author{Alejandro Jimeno-Pozo}
\thanks{These authors contributed equally.}
\affiliation{Fundaci\'{o}n IMDEA Nanociencia, C/ Faraday 9, Campus Cantoblanco, 28409 Madrid, Spain}
\author{Jose Angel Silva-Guill\'en}
\email{Corresponding author: joseangel.silva@imdea.org} 
\affiliation{Fundaci\'{o}n IMDEA Nanociencia, C/ Faraday 9, Campus Cantoblanco, 28409 Madrid, Spain}
\author{Francisco Guinea}
\email{Corresponding author: paco.guinea@imdea.org}
\affiliation{Fundaci\'{o}n IMDEA Nanociencia, C/ Faraday 9, Campus Cantoblanco, 28409 Madrid, Spain}
\affiliation{Donostia International Physics Center, Paseo Manuel de Lardiz\'{a}bal 4, 20018 San Sebasti\'an, Spain.}

\begin{abstract}
Rhombohedral graphene systems with different number of layers feature an abundance of correlated phases and superconducting states in experimental measurements with different doping and displacement fields.
Some of the superconducting pockets can emerge from - or close to - one of the correlated states.
Therefore, studying the phase diagram of the correlated phases for varying number of layers could be useful to interpret the experimental observations.
To achieve this, systematic Hartree-Fock calculations have been performed to build the phase diagram of rhombohedral (ABC-stacked) graphene for different number of layers, in the presence of long-range Coulomb interactions. 
By varying the external displacement field and carrier density, a cascade of metallic partially-isospin-polarized phases that spontaneously break spin and/or valley (flavor) symmetries is found. 
In addition, these states can present nematicity, stabilized by electron-electron interactions, exhibiting rich internal complexity. 
Polarized states are more stable for electron doping, and they are found for systems with up to 20 layers.
Moreover, the tunability of the phase diagram via substrate screening and spin-orbit coupling proximity effects has been studied.
Our results offer new insights into the role of correlations and symmetry breaking in graphitic systems which will motivate future experimental and theoretical works.
\end{abstract}

\maketitle
\section{Introduction}

Rhombohedral multilayer graphene (RMG) has recently emerged as a versatile platform for exploring strong electronic correlations and unconventional quantum phases. 
Experimental studies dealing with different multilayer devices have shown the tunability of correlated phenomena in rhombohedral graphene (RG) with doping and displacement fields.
Actually, several members of the RMG family have been found to exhibit a wide variety of correlated phases, including integer and fractional anomalous quantum Hall effects~\cite{Lu2023_FQHE, Zhou2024_FQHE}, Wigner crystallization~\cite{Tsui2024_wignerBBG}, and magnetic field/spin-orbit coupling (SOC)-induced superconductivity~\cite{Zhou2021SuperRTG,zhou2022isospin, zhang2023spin}, highlighting the rich interplay between symmetry, topology, and interactions in these systems.

In bilayer graphene, interaction-driven half-metal and other correlated states have been observed~\cite{zhou2022isospin, seiler2022quantum, barrera2022cascade, seiler2022quantum, barrera2022cascade, zhang2023spin, Seiler2024}, including isospin ordering and nematic transitions~\cite{Holleis2025, Seiler2025}.
Superconductivity has also been reported when applying a magnetic field which arises from a partially isospin-polarized state~\cite{zhou2022isospin}.
Moreover, by inducing SOC in the graphene layers, the superconducting critical temperature is increased~\cite{zhang2023spin} and the superconducting pocket can emerge from a nematic state~\cite{Holleis2025}.
Similar phase diagrams have been found in rhombohedral trilayer graphene with half and quarter-metal states~\cite{Zhou2021HalfMetRTG,Arp2024_IVC_SOC_RTG}. 
In this system, the superconducting pockets can arise from different normal states~\cite{Zhou2021SuperRTG}.

In higher-order rhombohedral stacks, such as four- to seven-layer graphene, signatures of broken-symmetry states and tunable anomalous Hall responses have been widely reported \cite{Ketal21, Liu2023, Han2023_5layer, Han2023_multiferroicity, Xie2024_6layer_hBN, Morissette2025_6layer, Zhou2024_7layer, Ding2025_7layer, Xiang2025_7layer}.
Furthermore, the emergence of chiral superconductivity in electron-doped tetralayer graphene, which develops close a quarter-metal phase, has been recently measured~\cite{Han_ChiralSC_tetralayer_2024}.

Interestingly, correlated phenomena such as electronic phase separation~\cite{NKSM18, Shi2020RTG},  and topological flat bands~\cite{Zhang2024_graphiteTopo, Xiao2025_graphiteTopo} have also been identified in the bulk limit. 
Nowadays, due to the advancements in the experimental techniques~\cite{yang2019stacking}, it is possible to grow clean samples of multilayer systems which allows to study the phase diagram of RMG with the number of layers as a tuning parameter as well~\cite{Shi2020RTG,Zhang2024_layerDependentCorr}.

On the theoretical side, several approaches have been employed to understand the origin of these plethora of correlated phases, particularly in the presence of external fields, quasi-flat bands, and spin-valley degeneracy lifting. 
Hartree-Fock calculations have provided insights into the phase diagrams of bilayer and trilayer graphene with and without Ising SOC~\cite{Xie2023_BBG_SOC, Koh2024, Wang2024, Zhumagulov2024, Huang2023_RTG, Koh2024_RTG}, while other works have analyzed Wigner crystal formation~\cite{AguilarMendez2025_BBG_wigner}, cascade transitions~\cite{Chichinadze2022, Friedlan2025}, and field-tunable symmetry breaking~\cite{ghazaryan2021unconventional, Ghazaryan2023Multilayers, Dong2023spin, Dong2023}. Alternative approaches for analyzing the phase diagram in  bilayer~\cite{Szabo2022_Bilayer, Mayrhofer2025_BBG} and trilayer~\cite{Das2024} graphene have also been applied.
Notably, the superconducting instabilities have also been thoroughly studied in bilayer, trilayer, and ABCA-stacked systems~\cite{JimenoPozo2023, Pantaleon2023ReviewSC, ZiyanLi2023, Chou_Intravalley_tetralayer_2024, Geier_isospin_tetralayer_2024, ParraMartinez2025_SC}, as well as the emergence of inter-valley coherent states and isospin ordering~\cite{Chatterjee2022, Huang2023_RTG, you2022_kohn}. 

Although substantial progress has been made, a comprehensive understanding of the interaction-driven ground states across the broader family of rhombohedral $N$-layer graphene is still lacking. 
In this work, we perform mean-field calculations within the self-consistent Hartree-Fock approximation to systematically explore the flavour-symmetry-broken phase diagram as a function of the external displacement field and carrier density in rhombohedral graphene with a varying number of layers ($N$).
We find a plethora of correlated phases where isospin polarized and partially isospin polarized states emerge in the phase diagram for a large number of doping densities and displacement fields. 
Nematic states, in which the symmetry of the Fermi surface is broken, also emerge within these polarized states in a systematic manner for all the number of layers considered.
Moreover, we also study in detail how the phase diagram of RG can be tuned by changing its environment, either by modifying the dielectric constant or by including spin-orbit coupling, which may be realized through substrates that have a strong SOC such as transition metal dichalcogenides, in our calculations. 
%we also study in detail how the dielectric environment can be employed to tune the phase diagram of RG, and assess the impact of incorporating spin–orbit coupling which may be realized through substrates that induce an effective SOC such as WSe$_{2}$. 
We observe that, by reducing the dielectric constant, the broken symmetry phases shift towards larger dopings and displacement fields. 
Conversely, new correlated phases are stabilized upon the inclusion of Ising SOC. 
Furthermore, we show how, although the dome of broken symmetry phases shrinks as the number of layers is increased, it survives for systems with more than 20 layers, implying that these could persist up to the graphite limit.
%We also study the effect of the dielectric environment as well as SOC in their phase diagram. \textcolor{orange}{Yo quitaría esta frase anterior y la de abajo que empieza for "Furthermore, ..." porque son repetitivas, y pondría aqui algo mezclando ambas.} 
%We find that there is a plethora \textcolor{orange}{quitar "that there is" y que quede solo "We find a plethora of ..."} of correlated phases where isospin polarized and partially isospin polarized states emerge in the phase diagram for a large number of doping densities and displacement fields.
%Nematic states, in which the symmetry of the Fermi surface is broken, also emerge within these polarized states. 
%Furthermore, we study in detail how the phase diagram of RG changes with the number of layers as well as how it can be tuned by changing the dielectric environment or by including spin-orbit coupling. 
%{\color{red} 
%We show how, although the dome becomes smaller as the number of layers is increased, it survives for systems with more than 20 layers.
%By changing the dielectric environment the states shift towards higher dopings and displacement fields and we find that if a substrate with strong SOC new phases are stabilized.}

The analysis of these metallic (mean-field) ground states has important implications for the realization of correlated phases in the rhombohedral multilayer graphene family. 
Moreover, as reported in experimental and theoretical works~\cite{zhang2023spin, JimenoPozo2023, Yang2024_SOC_SC_RG, Patterson2025, Zhang2025}, the superconducting phases of graphitic systems can arise from one of these phases. 
Therefore, analyzing in a fundamental way their phase diagram could allow for further understanding of the exotic superconducting states.
Our results provide a unifying framework for comprehending interaction-driven symmetry breaking in rhombohedral graphene systems and establish a foundation for further theoretical and experimental exploration.

\section{Results and discussion}\label{sec:results}

\subsection{General phase diagram}

We have studied the mean-field phase diagram (see Sec.~\ref{sec:methods} for a detailed description of the calculations) from bilayer to heptalayer rhombohedral stacked graphene, both for electron and hole doping under a perpendicular electric field.
One of the main results of our work is shown in the phase diagram of Fig.~\ref{fig:PD10}, where our Hartree-Fock calculations reveal a rich landscape of interaction-driven phases across the rhombohedral $N$-layer graphene family.
A prominent and recurring feature is the presence of isospin-polarized states (IPs) over a wide range of carrier densities and displacement fields. The IPs are characterized by a non-symmetrically filling of its isospin population $(n_{+\uparrow},n_{+\downarrow},n_{-\uparrow},n_{-\downarrow})$ in spin and valley degrees of freedom (see Table ~\ref{tab:brokenSym_states} for a description of the different IPs).

\begin{figure}[t]
\centering
\includegraphics[width=0.44\textwidth]{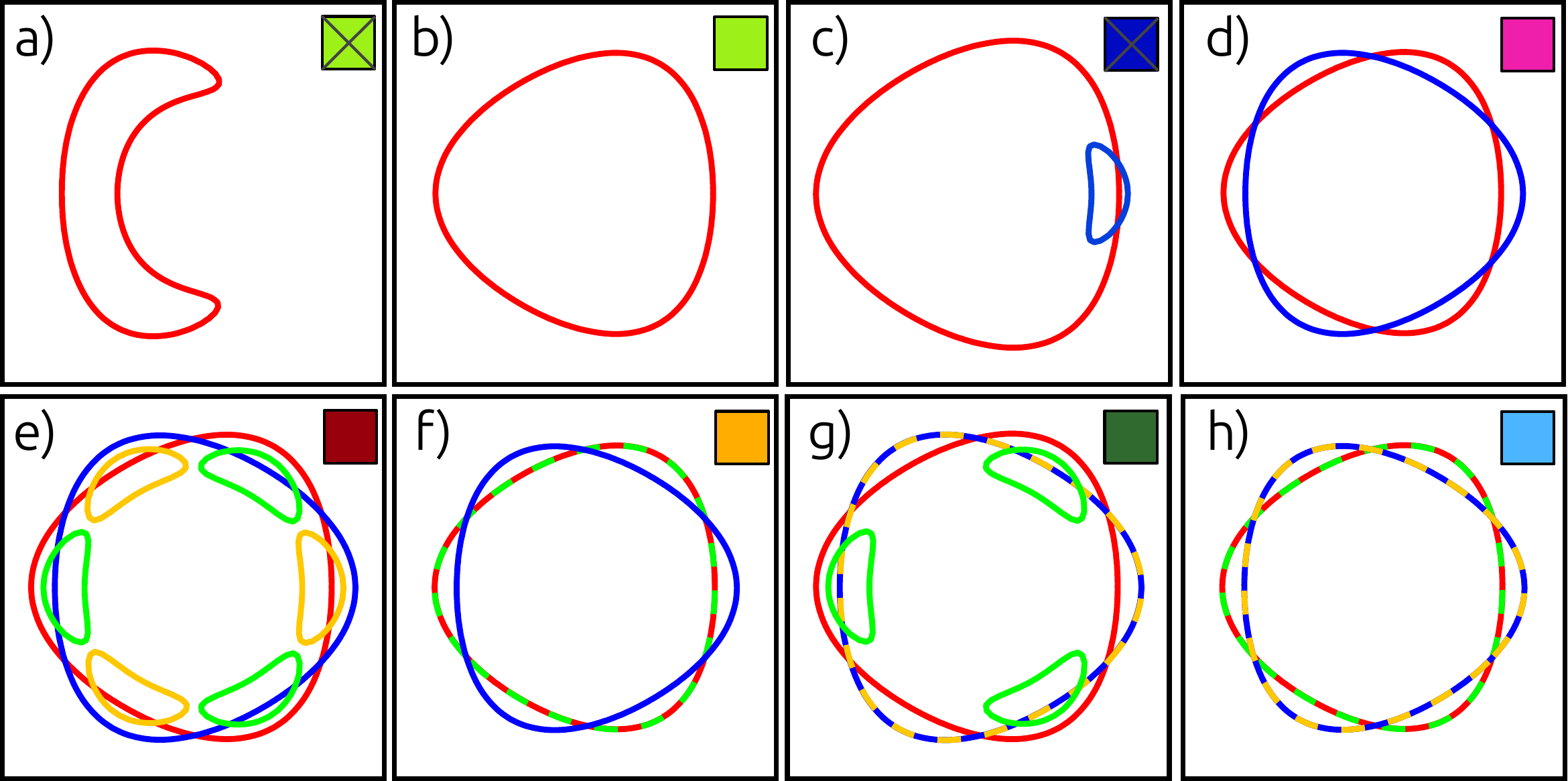}
\caption{Isospin resolved Fermi surfaces for different polarized states. Data taken from $N=5$ at $D=220$ meV and $\epsilon=10$.}
\label{fig:FS}
\end{figure}

For electron doping, these IPs are present in all the systems that we studied, revealing that electronic interactions favor isospin ordering across the entire multilayer family.
The IPs, independently of the number of stacked layers, follow an ordered cascade of polarization states. 
Taking the trilayer case ($N=3$) as an example, for $D$ = 200 meV, by varying $n_e$ we go from a (isospin-filled) quarter metal state to the full symmetric state passing through a half and a three-quarter metal. 
In addition to these progressively-filling ordering, a \textit{partially} isospin-polarized state (PIPs) emerges. 
We label PIPs with two subidexes, $i,j$, corresponding to the number of fully filled and partially filled flavors respectively.
For a larger number of layers, we have identified several PIPs which mediate between IPs (all possible PIPs are listed in Table \ref{tab:brokenSym_states}).
This family of states naturally arises mediating between IPs as the system energetically prefers to start partially filling a new flavor rather than highly populating  some flavors over the empty ones. 
In PIPs, the polarized bands are not identically populated and the Fermi surfaces (FS) are different for each isospin.
This can be seen in Figs.~\ref{fig:FS}e and g, where the FS is composed of a fully filled isospin flavor and other partially filled isospin flavor. 
This behavior is expected for intermediate phases that interpolate between fully polarized and unpolarized configurations, indicating that PIP states represent intermediate configurations that relieve the energetic cost of abrupt transitions between qualitatively different Fermi surface topologies.
These states can host spin or valley partial polarization which makes them of great interest as electronic correlations could be enhanced and spintronics and valleytronics are highly sensitive to small doping and external electric field changes.

Another key feature that we observe in the phase diagrams is the emergence of nematicity (see the regions with crosses in the different IPs in Fig.~\ref{fig:PD10}), that is, the breaking of $C_3$ symmetry.
This breaking is clear by plotting the FS of the state. 
For example, in Fig.~\ref{fig:FS}a the FS of the nematic QM state is not $C_3$ symmetric while that of the non-nematic state, Fig.~\ref{fig:FS}b, is.
Nematicity in multilayer graphene systems has been previously reported mainly for IPs~\cite{ParraMartinez2025_SC,Koh2024,Koh2024_RTG}.
Here, as shown in Fig.~\ref{fig:PD10}, nematicity arises at low electronic fillings and moderate to high displacement fields primarily and is stable in large regions of the QM and PIP$_{11}$.
Interestingly, analyzing carefully the nematic states we found that not all of them display the same order of nematicity.
In fact, calculating what we define as the nematic order parameter (see Sec.~\ref{sec:methods}C) we detect that in some PIPs it shows a lower value than those present in the QM phase.
This can be understood by looking at the Fermi surface shown in Fig.~\ref{fig:FS}c where we can see how the fully polarized isospin is filled enough so that it displays a symmetric rounded FS (red) while the partially polarized isospin (blue) undergoes a $C_3$ symmetry-breaking as the QM does at lower fillings.
Therefore, this type of phases exhibit \textit{partial}-nematicity. 

As the number of graphene layers increases, the location of the broken-symmetry phases systematically shifts towards larger carrier densities and weaker displacement fields.
This trend may be partially understood by noting that the non-interacting density of states (DOS) becomes more asymmetric with increasing $N$, showing a notable enhancement on the hole side, see Sec. SI in the Supporting Information (SI)~\cite{SM}. 
The increased DOS at the vHs enhances the screening, resulting in a shift on the stability regions of symmetry-broken states. 
As a general feature, we find that increasing the number of layers pushes the boundaries of all isospin-symmetry-breaking phases toward higher carrier densities.

Now we turn our attention to the effect of hole-doping in $N$ layer RG. 
For $N=2$ and 3, as in the case of electron doping, we find a cascade of IPs and PIPs, including some nematic states. 
Interestingly, for large displacement fields the most stable phase is the fully symmetric. 
These results agree with the cascade of phases observed in Ref.~\cite{zhou2022isospin} where they identify several PIPs and a quarter metal state.
As the number of layers is increased, the correlated states are suppressed and the normal metal extends over all the phase diagram for $N=7$, which is in stark contrast to the behaviour in the electron side.
For those number of layers where the distinct phases emerge, the states shift towards larger values of doping and lower displacement fields as in the case of electron doping.

\subsection{Tuning the phase diagram: Dielectric screening and Ising spin-orbit coupling}

Experimentally, multilayer RG samples are placed on top of (or sandwiched between) different substrates. 
There can be two principal consequences when changing the substrate: a change in the dielectric constant or, if the material has a strong SOC it can induce it in the graphene layers. 
In this section, we analyze which is the effect of these changes separately. 
We do not consider internal screening of the displacement field by the graphene layers. The Thomas-Fermi screening length of rhombohedral graphite is expected to be large~\cite{Kotov2012}, because of its low density of states at the Fermi level. On the other hand, the electronic structure of rhombohedral stacks leads to charge accumulation at the top and bottom layers, but the doping levels considered here will lead to  electric fields lower, even if the charge is mostly concentrated at the top and bottom layers.

\begin{figure*}[h!]
\centering
\includegraphics[width=1.0\textwidth]%{Figures/HF10D.pdf}
{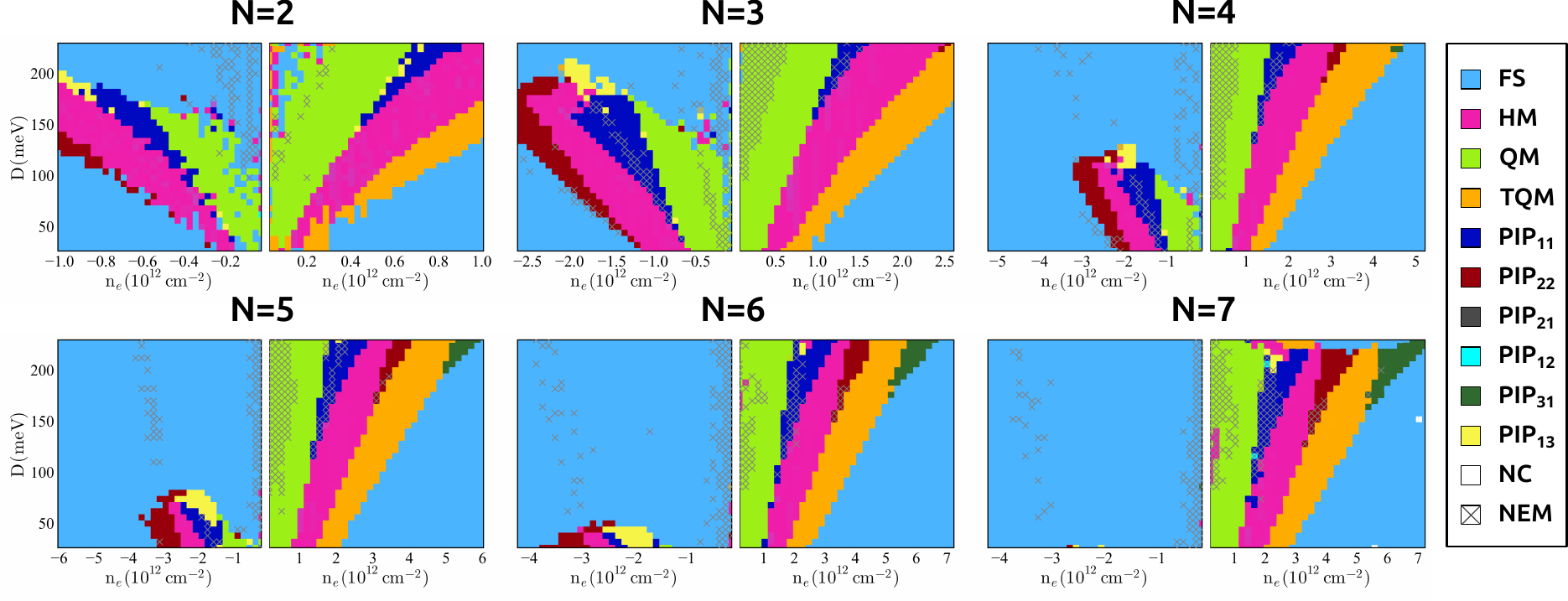}
\caption{Hartree-Fock phase diagrams for rhombohedral multilayer graphene with different number of layers $N$. We use $\epsilon=10$. The different phases are depicted in different colors and are described in Table~\ref{tab:brokenSym_states}.}
\label{fig:PD10}
\end{figure*}

\begin{figure*}[t]
\centering
\includegraphics[width=1\textwidth]{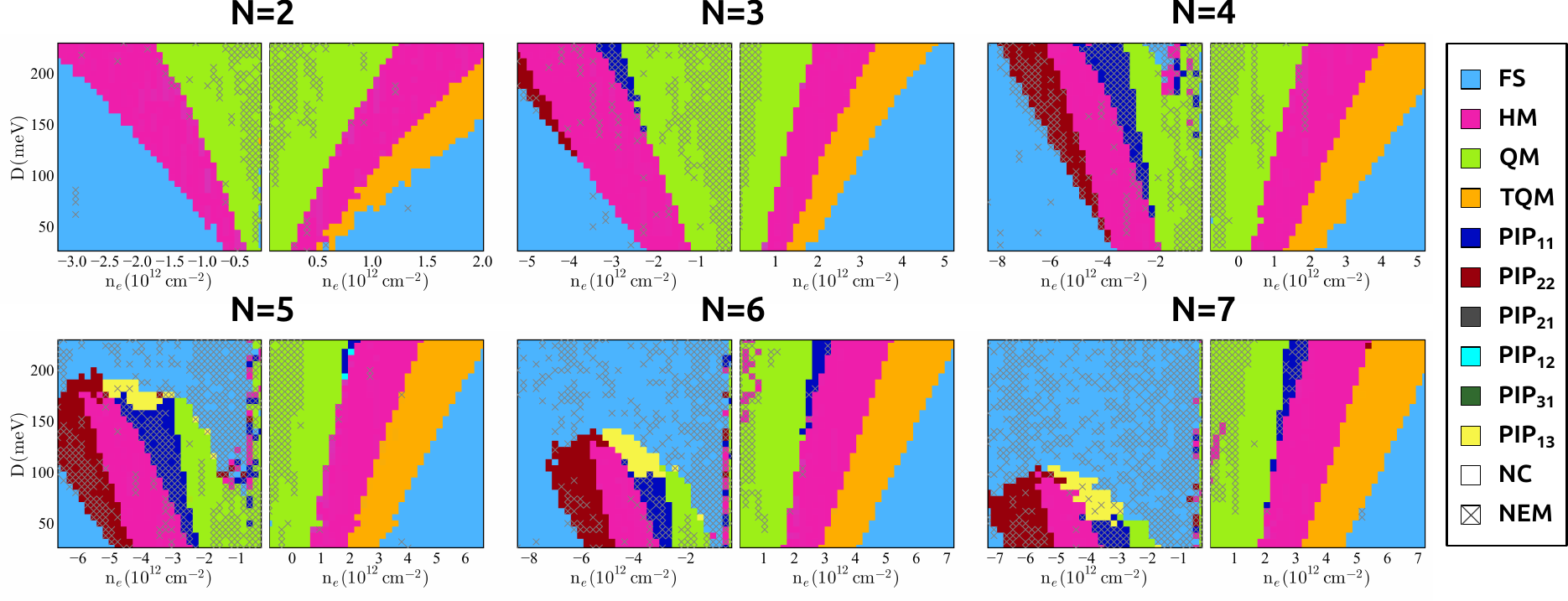}
\caption{Hartree-Fock phase diagrams for rhombohedral multilayer graphene with different number of layers $N$. We use $\epsilon=4$. The different phases are depicted in different colors and are described in Table~\ref{tab:brokenSym_states}.}
\label{fig:PD4}
\end{figure*}

We begin analyzing the change of the dielectric environment of the graphene layers.
To calculate the phase diagram shown Fig.~\ref{fig:PD4} we use a dielectric constant of $\epsilon=4$. % (we used $\epsilon=10$ in Fig.~\ref{fig:PD10}).
We can see how the main feature found in Fig.~\ref{fig:PD10} for $\epsilon=10$ is preserved: there is still a cascade of different states with IPs and PIPs.
Nevertheless, there are differences when comparing both phase diagrams.
For example, the area of the correlated phases expands over a larger area of the phase diagram for a smaller $\epsilon$.
In fact, the states shift to larger values of electron density and displacement field leading to a reduced (enhanced) prevalence of the PIP phases in Fig.~\ref{fig:PD4} compared to Fig.~\ref{fig:PD10} in electron (hole) dopings. 
Moreover, the dome of correlated phases in the hole side survives to the addition of layers, being still present for $N=7$. 
%{\color{red} This is expected since a reduced dielectric constant enhances the Coulomb interaction, strengthening the tendency toward interaction-driven symmetry breaking. However, as the number of layers increases, internal screening becomes more effective, which in turn suppresses these interactions and leads to the eventual disappearance of correlated phases.}
Therefore, our results from Figs.~\ref{fig:PD10} and~\ref{fig:PD4} show that PIPs can be highly tunable by increasing the number of layers or the screening.

\begin{figure*}[t]
\centering
\includegraphics[width=1\textwidth]{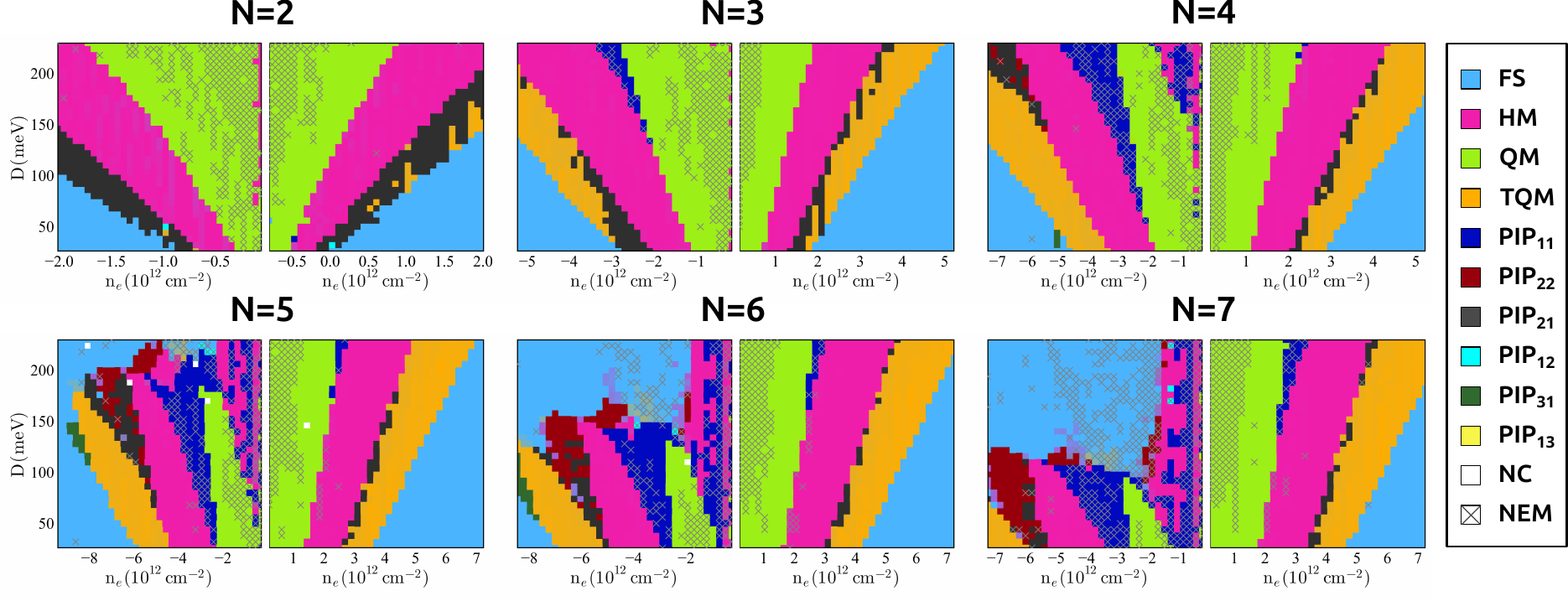}
\caption{Hartree-Fock phase diagrams for rhombohedral multilayer graphene with different number of layers $N$. We use $\epsilon=10$ and include SOC with $\lambda_I=2$ meV. The different phases are depicted in different colors and are described in Table~\ref{tab:brokenSym_states}.} 
\label{fig:PDsoc}
\end{figure*}

We now examine the phase diagram when SOC is induced in the graphene layers. 
These calculations were performed using a value of $\lambda_I=2$ meV and $\epsilon=4$. 
As shown in Fig.~\ref{fig:PDsoc}, the most distinctive characteristic is the stabilization of the PIP$_{21}$. 
This state arises in competition with the three-quarter metal (TQM) which, as the number of layers increases, ends up energetically beating the PIP$_{21}$. 
The emergence of the TQM on the hole side is also a unique feature result of Ising SOC as it was not the ground state for negative fillings in Figs.~\ref{fig:PD10},~\ref{fig:PD4}. 
We also highlight the presence of a non-ordered phase competition between HM and PIP$_{11}$ at negative fillings and $N>4$ layers. 
Ising SOC lifts off the degeneracy present in the half metal states between valley-half metal (VHM) and spin-half metals and favours a spin-valley locked half metal with a polarization vector $\vec{P}=(\alpha,0,0,\alpha)$.

\subsection{Bulk limit}

\begin{center}
\begin{figure}[t]
\includegraphics[width=0.45\textwidth]{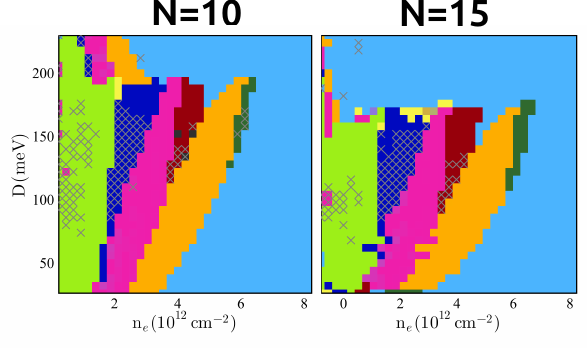}
\caption{Hartree Fock phase diagrams for $N=10$ and 15. We use $\epsilon=10$. The different phases are depicted in different colors and are described in Table~\ref{tab:brokenSym_states}.} 
\label{fig:PDNlarge}
\end{figure}
\end{center}

As we have stated in the Introduction, nowadays it is possible to fabricate, in a very clean manner, devices with a specific number of layers, $N$.
Furthermore, topological phases have been found in bulk samples.
Here we try to elucidate if the correlated phases survive up to the bulk limit or, on the contrary, if there is a metallic limit where the phase diagram is governed by the full metal state.
To do so, we have performed calculations on systems with $N=10$ and 15 for electron doping and $\epsilon=10$. 
The results are shown in Fig.~\ref{fig:PDNlarge}.
In both cases, the dome of states is still present and the phases are the same as in the case of $N=7$.
Nevertheless, for $N=10$ we observe that for intermediate values of $n_e$ and large values of $D$, the full symmetric phase is the most stable.
Interestingly, as we increase the number of layers up to 15, the full symmetric phase completely dominates the phase diagram for large values of $D$ and the dome has shrank significantly.
Another feature that is worth mentioning is the appearance of the PIP$_{13}$ which is beginning to be more stable than PIP$_{11}$.
We have also performed calculations for several points of the phase diagram for $N=20$ and have identified that the correlated phases are still present, although for low doping values and intermediate values of displacement field the FS phase is being stabilized. 
Moreover, for low values of $D$ this phase is also winning some ground.

\section{Conclusions}

Recent works on rhombohedral graphene devices have revealed a rich and intricate landscape of physical phenomena.
In this work, we have studied the phase diagram by performing Hartree-Fock calculations of RG for varying number of layers.
We have found that the emergence of broken-isospin-symmetry phases driven by electron interactions seems to be ubiquitous to the RG family. 
Actually, although at first glance we could think that the transition from a two-dimensional system to the bulk is trivial, our calculations show that it is not the case and that even for $N>20$ there are still correlated phases present if RG is electron-doped.
For the hole-doped case, the trivial metal state extends over all the phase diagram for $N>7$ if $\epsilon=10.$
We have also analyzed the tunability of the phase diagram since experimentally different substrates can be used to support the RG samples.
If the dielectric constant is reduced, the correlated states extend over larger values of doping and electric field. 
In fact, for this type of environment, the dome in the hole-doped side of the phase diagram is still present even for $N>7$.
Moreover, by inducing spin-orbit coupling in the graphene layers mimicking the effect of a substrate with a strong SOC such as a transition metal dichalcogenide, new correlated states emerge, but the overall shape of the dome remains the same. 
Our work highlights that correlations are at the origin of the rich phase diagram featured in graphitic systems and could be useful in the interpretation both for the exotic phases and the superconducting states found experimentally in these systems. 

\section{Methods}\label{sec:methods}

\subsection{Non-interacting model}

We start by defining our non-interacting Hamiltonian. 
We employ a continuum model described in Refs.~\cite{Min2008, Min2011} that includes two electrons per layer in the sublattice (A,B) space around the Fermi level. 
The single-electron Hamiltonian reads $\hat{H}_0=\sum_{\mathbf{k}\alpha\beta}\hat{c}_{\alpha,\mathbf{k}}^{\dagger}h_{\alpha,\beta}(\mathbf{k})\hat{c}_{\beta,\mathbf{k}}$ where the operator $\hat{c}_{\alpha,\mathbf{k}}^{\dagger}$ ($\hat{c}_{\alpha,\mathbf{k}}$) creates (annihilates) an electron with momentum $\mathbf{k}=(k_x,k_y)$ measured with respect to the Dirac cone valley and quantum numbers $\alpha=(\sigma,\xi)$ running on the tensor-product space of spin ($\sigma=\uparrow,\downarrow$) and valley ($\xi=+,-$) degrees of freedom.
The matrix elements of $h_{\alpha,\beta}(\mathbf{k})$ for a given flavour are given by,
\begin{equation}
\label{eq:main_hamil}
     h_{\alpha,\beta}(\mathbf{k}) = \begin{bmatrix}
\mathcal{L}+\mathcal{U}_1 & \mathcal{V} & \mathcal{W} & 0 & \cdots & 0 \\
 & \mathcal{L} & \mathcal{V} & \mathcal{W} & \ddots & \vdots \\
 &  & \mathcal{L} & \mathcal{V} & \ddots & \vdots \\
 &  &  & \ddots & \ddots & \mathcal{W} \\
 &  &  &  & \ddots & \mathcal{V} \\
 &  &  &  &  & \mathcal{L}+\mathcal{U}_N
\end{bmatrix},
\end{equation}
where
\begin{subequations}
    \begin{equation}
        \mathcal{L} = \mqty(0 & v_{0}\pi^{\dagger} \\ 0 & 0),
    \end{equation}
    \begin{equation}
        \mathcal{V} = \mqty(v_{4}\pi^{\dagger} & v_{3}\pi \\ \gamma_{1} & v_{4}\pi^{\dagger}),
    \end{equation}
    \begin{equation}
        \mathcal{W} = \mqty(0 & \gamma_{2}/2 \\ 0 & 0),
    \end{equation}
\end{subequations}
with $\pi=\xi k_x+ik_y$ and $v_{i}=\frac{\sqrt{3}}{2}a\gamma_{i}$. 
For simplicity, we assume that an external perpendicular displacement field induces a potential difference, $\text{D}$, between the 1-$st$ and the $N$-$th$ layer. The parameter $\delta$ accounts for the on-site energy difference between sites A$_1$ and B$_N$ with respect to the high energy sites, with contributions $ \mathcal{U}_1 = \text{D}\mathbb{I} +\delta(\mathbb{I}+\sigma_z)/2$ for the 1st layer and $ \mathcal{U}_N = \text{D}\mathbb{I} +\delta(\mathbb{I}-\sigma_z)/2$ for the $N$ layer.

In the calculations, the continuum model is defined in a hexagonal grid centered at the Dirac point up to a momentum cut-off given by $\Lambda_{c} = 0.06 K_{D}$, where $K_{D}=\frac{4\pi}{3 a}(1,0)$ is the position of the Dirac point, corresponding to one corner of the Brillouin zone.
The numerical values used for the continuum model parameters are listed in Tab.~\ref{tab:hopping_params} and taken from existing literature~\cite{Zhou2021HalfMetRTG, Zibrov2018GullyTG}.

\subsection{Interacting model}
In this section, we define the interacting hamiltonian and summarize the self-consistent Hartree-Fock computational framework. We start from the two-particle Coulomb interaction: %\ref{SM: Electronic}
\begin{equation}
\label{eq:four-fermion Coulomb}
    \hat{\mathcal{V}}_\mathcal{C} = \frac{1}{2A}  \sum_{\substack{\mathbf{k}, \mathbf{p}, \mathbf{q}}}\sum_{\substack{\alpha, \beta}} \mathcal{V}_0(\mathbf{q}) \hat{c}^\dagger_{\alpha, \mathbf{k}  + \mathbf{q} }\hat{c}^\dagger_{\beta, \mathbf{p} - \mathbf{q}}\hat{c}_{\beta, \mathbf{p}}\hat{c}_{\alpha, \mathbf{k} },
\end{equation}
where $\hat{c}$ and $\hat{c}^\dagger$ denote annihilation and creation fermionic operators respectively, and $\mathcal{V}_{0}(\mathbf{q}) = \frac{2\pi e^2}{\epsilon |\mathbf{q}| }\tanh{\left(d_{g} |\mathbf{q}|\right)} $ is the double-gated screened Coulomb potential 
with $e$ being the electron charge, $\epsilon$ the dielectric constant associated with encapsulating the system, and $d_{g}$ the distance to the gates, which we choose to be 200 nm. The Coulomb interaction, $\hat{\mathcal{V}}_\mathcal{C}$, can be decomposed into two-fermion terms via mean-field approximation, that is, the Hartree and Fock contributions: \\
\begin{equation}
    \begin{split}
        \hat{\mathcal{V}}_C &\approx \sum_{ \substack{ \mathbf{k} ,\alpha}}  \left[ \mathcal{V}_0\left( \mathbf{0}\right)\sum_{\mathbf{p} , \beta}n_{\beta\beta}(\mathbf{p}) \right]\hat{c}^\dagger_{\alpha, \mathbf{k} }\hat{c}_{\alpha, \mathbf{k} }\\
& - \sum_{\substack{ \mathbf{k} ,\alpha, \alpha' }}  \left[\sum_{ \mathbf{p} }\mathcal{V}_0\left( \mathbf{p}- \mathbf{k} \right) n_{\alpha\beta}(\mathbf{p}) \right]\hat{c}^\dagger_{\alpha, \mathbf{k}} \hat{c}_{\beta, \mathbf{k}},
    \end{split}
\end{equation}
being $n_{\alpha\beta}(\mathbf{q})=\frac{1}{A}\left\langle \hat{c}^{\dagger}_{\beta,\mathbf{q}}\hat{c}_{\alpha,\mathbf{q}}\right\rangle$ the density matrix. Consequently, the resulting Hamiltonian can be solved self-consistently starting from different initial ansatzs. 
We design a general framework to study polarized and partially polarized states in $N$-layers graphene. 
In addition to that, we delve into the breaking of $C_3$ symmetry, which results in a nematic phase.
In Table ~\ref{tab:brokenSym_states}, we list the broken symmetry phases studied in this work. 
For each value of electronic density $n_e$ and perpendicular electric field $D$, we perform a self-consistent Hartree-Fock calculation for each possible state until convergence is reached for a tolerance lower than $10^{-6}$. 
The lowest-energy solution among the candidates is chosen as the ground state. 
The Hartree-Fock calculations are carried out in the hexagonal grid defined by $\Lambda_{c}$ containing 7651 $k$-points. 
Each of our phase diagrams include 1225 points ($35\times35$).
All calculations are performed at zero temperature.

We characterize the ground state by the polarization vector, that is, the electronic population for each isospin:
\begin{equation}
\vec{P}=(n_{+\uparrow},n_{+\downarrow},n_{-\uparrow},n_{-\downarrow})
\end{equation}
\begin{equation}
    n_{\xi,\sigma}=\frac{1}{    N   _k}\sum_{n,k}|\psi_{nk}^{(\xi\sigma)}|^2\Theta(\varepsilon_{nk}-\mu)
\end{equation}
We distinguish two primary groups of solutions: Isospin Polarized states (IPs) and Partially Isospin Polarized states (PIPs). 
The first group, IPs, consists of states where the system equally fills different flavours. While, for PIPs the system fills multiple flavours non-symmetrically. These PIPs result in different Fermi surfaces between isospin (beyond valley mirroring) which could be experimentally resolved.
We label the PIPs, by two subindexes PIP$_{i,j}$ where $i$ denote the number of fully ocuppied isospins and $j$ accounting for the fractionally filled isospins.

\subsection{Nematic order parameter}

The non-interacting Hamiltonian from the continuum model in Eq.~\ref{eq:main_hamil} has a $C_3$ symmetry inherited from the point group of multilayer graphene. 
However, it is possible to generate a seed of the ansatz that breaks this symmetry, which would allow us to search for nematic states in the system. 
To do so, we engineer all possible isospin-polarized-nematic states that are listed in Tab.~\ref{tab:brokenSym_states}. 
We quantify the breaking of the $C_3$ symmetry by calculating the following order parameter:
\begin{equation}
\phi=\sum_{k,n}(n_{n,k}-n_{n,C_{3k}})^2
\end{equation}
Nematicity may appear in all the polarized isospins such as a half-metals where both populated bands show a $C_3$ symmetry breaking. 
Nonetheless, one of the polarized isospins may undergo a $C_3$ symmetry breaking while other polarized isospins still show $C_3$ symmetry. 
This feature is particularly favored in partially polarized states where the isospin imbalance results in a partial-nematicity.

\subsection{Ising spin orbit coupling}

The inclusion of Ising spin-orbit coupling (SOC) is motivated by proximity effects arising when rhombohedral multilayer graphene is placed on substrates with strong intrinsic SOC, such as transition metal dichalcogenides (e.g., WSe$_2$). These substrates induce a valley-dependent spin splitting in the graphene layers, leading to an effective Ising SOC. 

In the continuum approximation, i.e. at low energies near the $K$ and $K'$ points, this SOC is captured by the term
\begin{equation}
    H_{\text{I}} = \xi \sigma \lambda_{\text{I}} \mathbb{I},
\end{equation}
This term preserves time-reversal symmetry but lifts partially the isospin degeneracy. 

In our Hartree-Fock framework, $H_{\text{I}}$ is incorporated into the single-particle Hamiltonian before the self-consistent procedure. As a result, the phase diagram becomes sensitive to Ising SOC, with new isospin-polarized phases emerging at the boundaries of PIPs phases.

\section{Acknowledgements}
We acknowledge \foreignlanguage{vietnamese}{Võ Tiến Phong} for illuminating discussions. 
IMDEA Nanociencia acknowledges support from the ‘Severo Ochoa’ Programme for Centres of Excellence in R\&D (CEX2020-001039-S/AEI/10.13039/501100011033). 
We acknowledge support from NOVMOMAT, project PID2022-142162NB-I00 funded by MICIU/AEI/10.13039/501100011033 and by FEDER, UE as well as financial support through the (MAD2D-CM)-MRR MATERIALES AVANZADOS-IMDEA-NC.
J.A. S.-G. has received financial support through the ``Ram\'on y Cajal'' Fellowship program, grant RYC2023-044383-I financed by MICIU/AEI/10.13039/501100011033 and FSE+.
G.P.-M. is supported by Comunidad de Madrid through the PIPF2022 programme (grant number PIPF-2022TEC-26326). 
%A.J.P acknowledges partial support by grant NSF PHY-2309135 to the Kavli Institute for Theoretical Physics (KITP).  
We thankfully acknowledge RES resources provided by BSC in MareNostrum5 to FI-2025-1-0032 and by CIEMAT in Xula to FI-2025-2-0024. 

\section{Conflict of Interest}
The authors declare no conflict of interest.

\section{Data Availability Statement}
The data that support the findings of this study are available from the corresponding author upon reasonable request.

\section{Keywords}
Rhombohedral multilayer graphene, broken-symmetry phases, correlated electron systems, spin-valley polarization, exchange interactions.

\bibliography{bibliography}

\clearpage
\begin{table}[h!]
\centering
\renewcommand{\arraystretch}{1.25}

\begin{tabular}{|m{1.7cm} m{1cm} m{2cm} m{1.5cm} m{1.2cm}|}
\hline%\hline
\parbox[c][0.8cm][c]{1.7cm}{\centering \textbf{Name}} &
\parbox[c][0.8cm][c]{1cm}{\centering \textbf{Symbol}} &
\parbox[c][0.8cm][c]{2cm}{\centering \textbf{$\vec{P}$}} &
\parbox[c][0.8cm][c]{1.5cm}{\centering \textbf{Fermi surface}} &
\parbox[c][0.8cm][c]{1.2cm}{\centering \textbf{Legend}} \\
\hline \hline
\multicolumn{5}{|c|}{\parbox[c][0.6cm][c]{7.5cm}{\centering \textbf{Isospin polarized states}}} \\

\hline \hline
\parbox[c][1.1cm][c]{1.7cm}{\centering Fully symmetric} &
\parbox[c][1.1cm][c]{1cm}{\centering FS} &
\parbox[c][1.1cm][c]{2cm}{\centering $\left(\alpha,\alpha,\alpha,\alpha\right)$} &
\parbox[c][1.1cm][c]{1.5cm}{\centering \includegraphics[width=1cm]{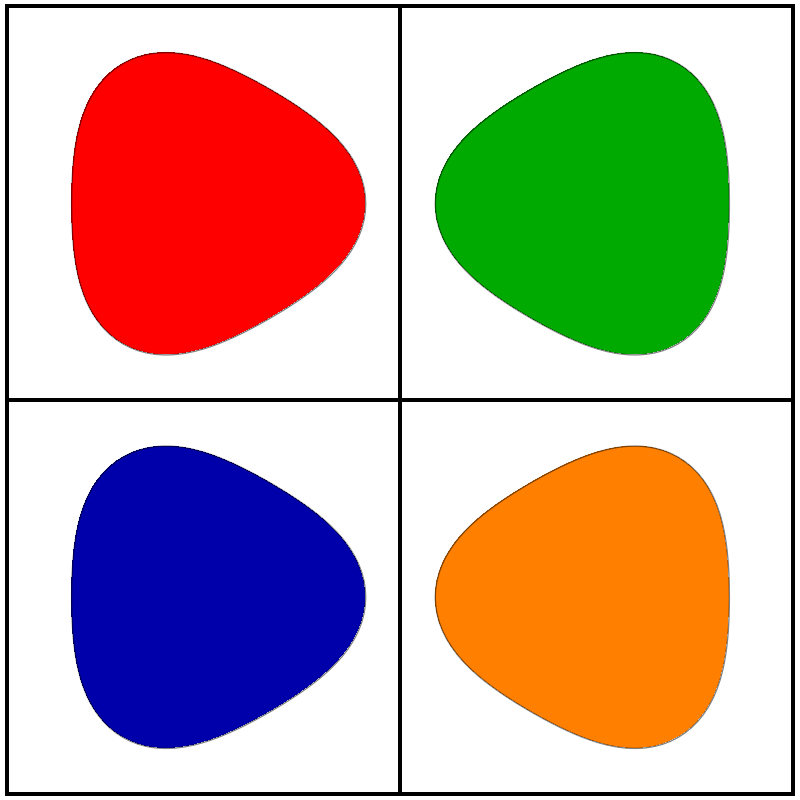}} &
\parbox[c][1.1cm][c]{1.2cm}{\centering \includegraphics[width=0.4cm]{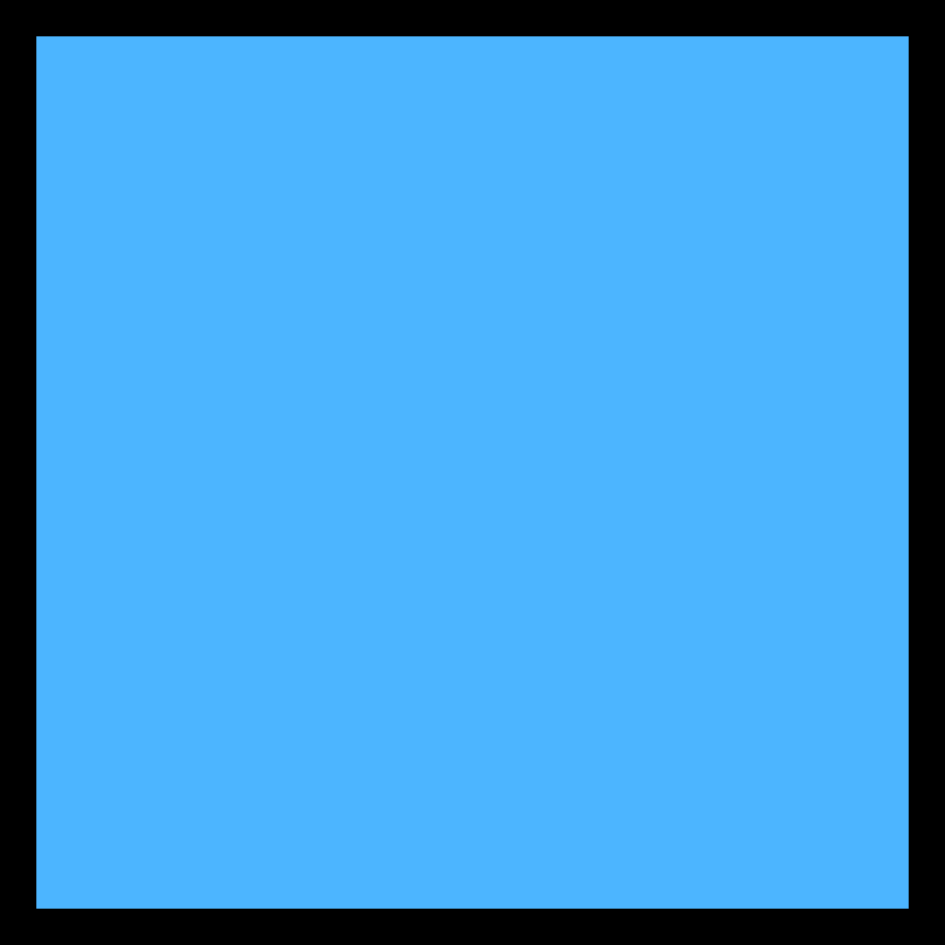}}\\
\hline

\parbox[c][1.1cm][c]{1.7cm}{\centering Half metal} &
\parbox[c][1.1cm][c]{1cm}{\centering HM} &
\parbox[c][1.1cm][c]{2cm}{\centering $\left(\alpha,0,\alpha,0\right)$} &
\parbox[c][1.1cm][c]{1.5cm}{\centering \includegraphics[width=1cm]{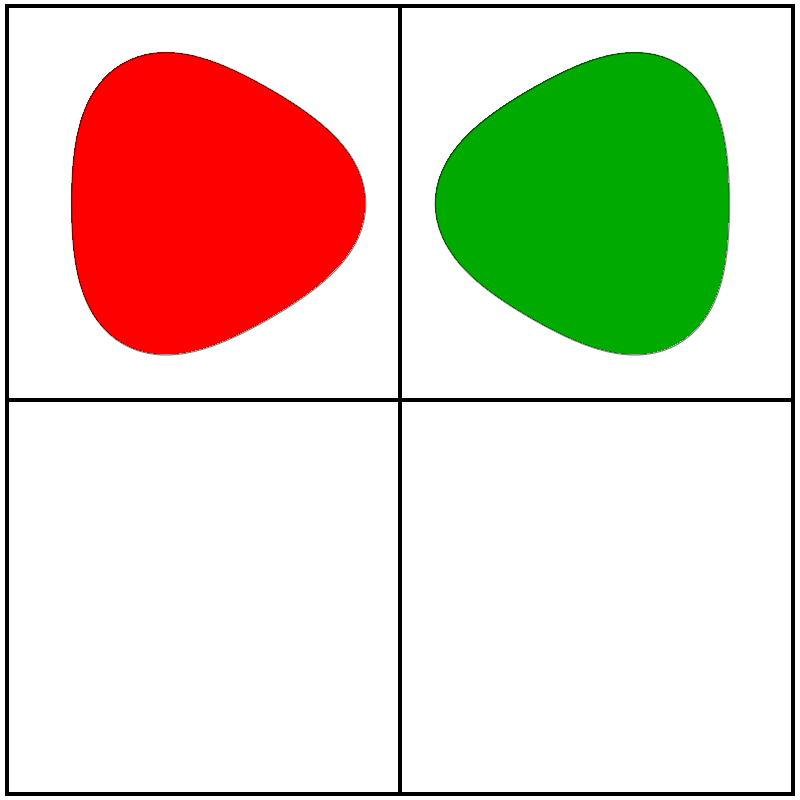}} &
\parbox[c][1.1cm][c]{1.2cm}{\centering \includegraphics[width=0.4cm]{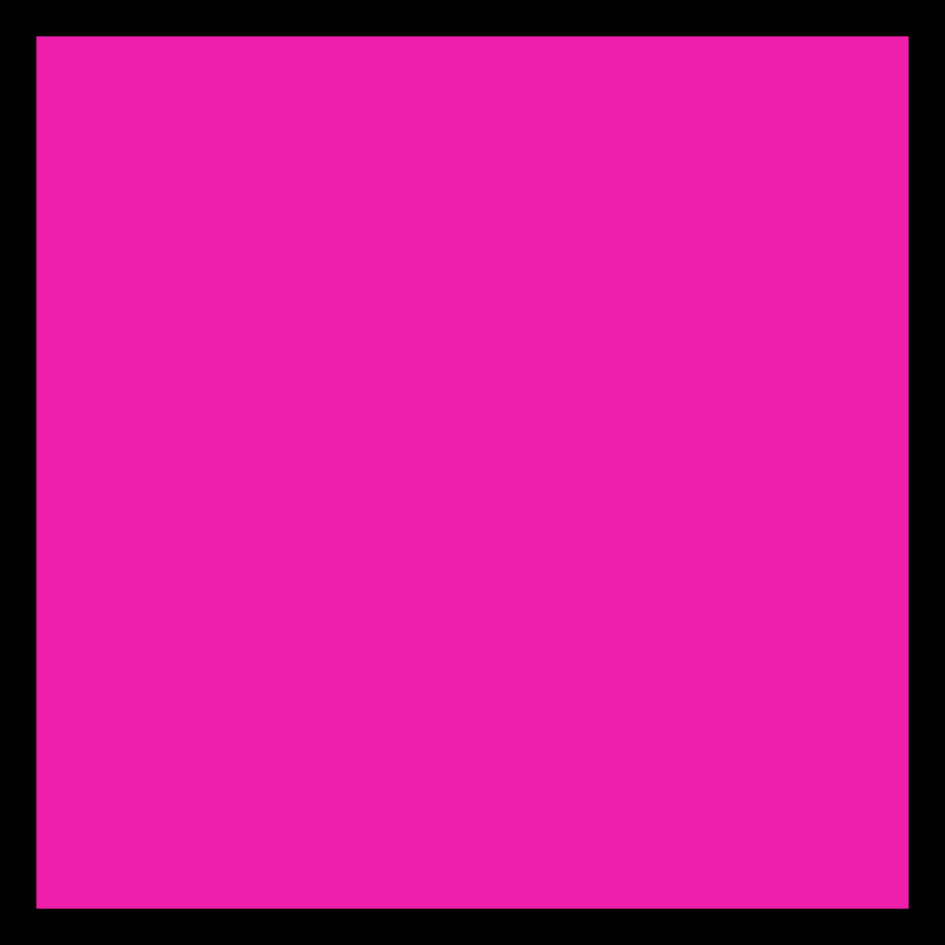}} \\
\hline

\parbox[c][1.1cm][c]{1.7cm}{\centering Quarter metal} &
\parbox[c][1.1cm][c]{1cm}{\centering QM} &
\parbox[c][1.1cm][c]{2cm}{\centering $\left(\alpha, 0, 0, 0\right)$} &
\parbox[c][1.1cm][c]{1.5cm}{\centering \includegraphics[width=1cm]{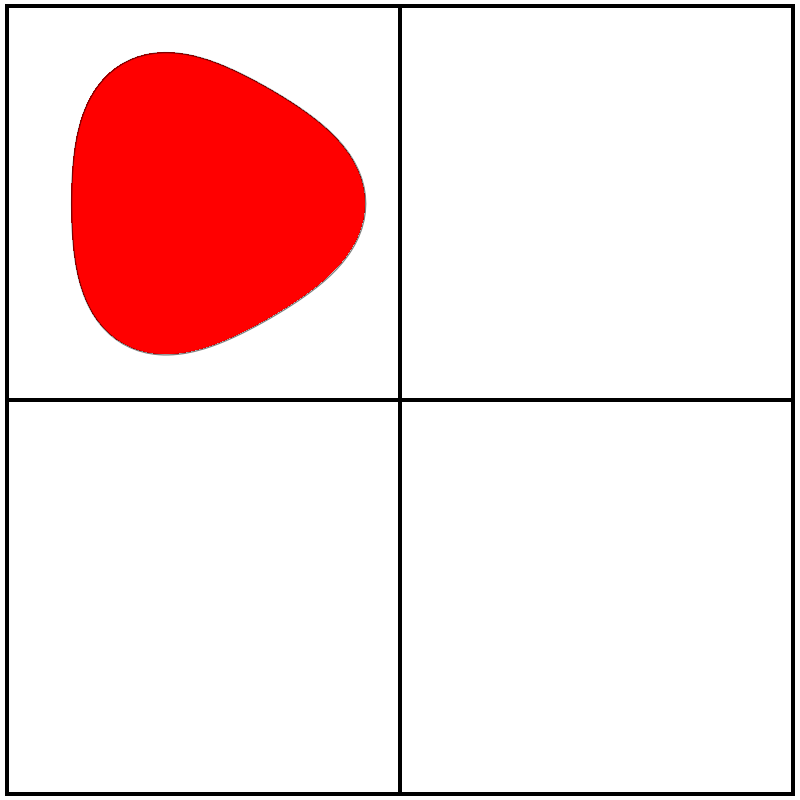}} &
\parbox[c][1.1cm][c]{1.2cm}{\centering \includegraphics[width=0.4cm]{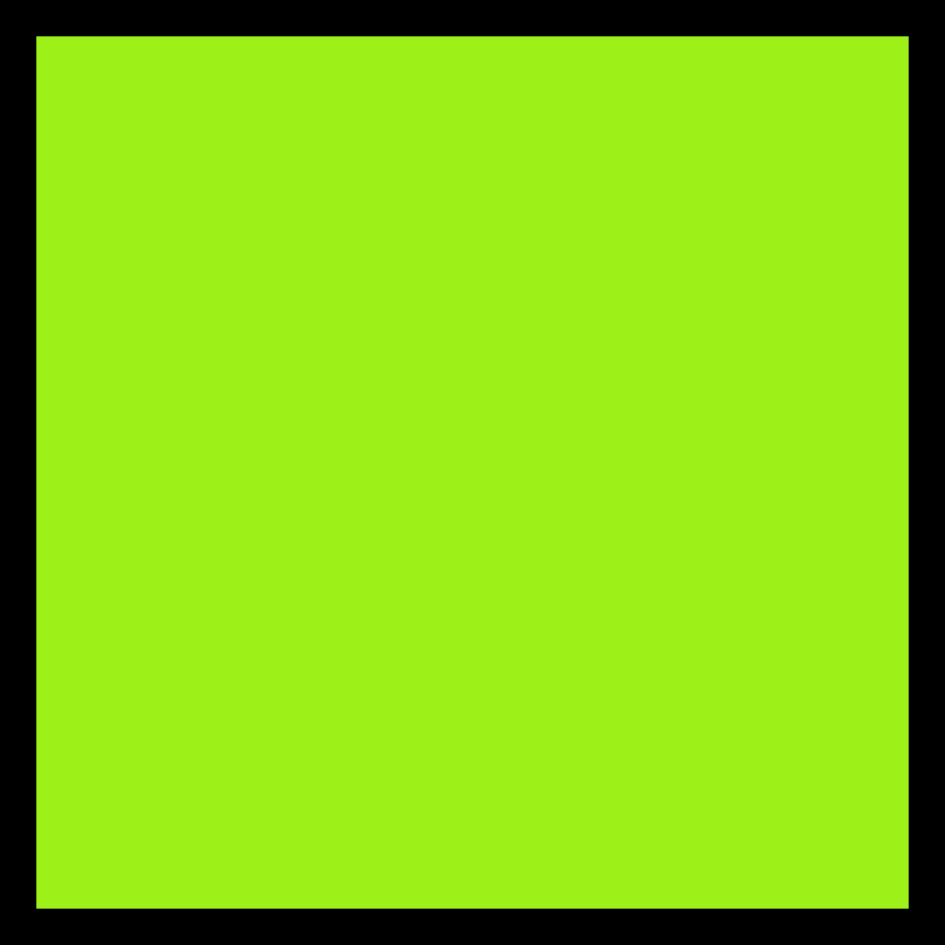}} \\
\hline

\parbox[c][1.4cm][c]{1.7cm}{\centering Three-quarter metal} &
\parbox[c][1.4cm][c]{1cm}{\centering TQM} &
\parbox[c][1.4cm][c]{2cm}{\centering $\left(\alpha, \alpha, \alpha, 0\right)$} &
\parbox[c][1.4cm][c]{1.5cm}{\centering \includegraphics[width=1cm]{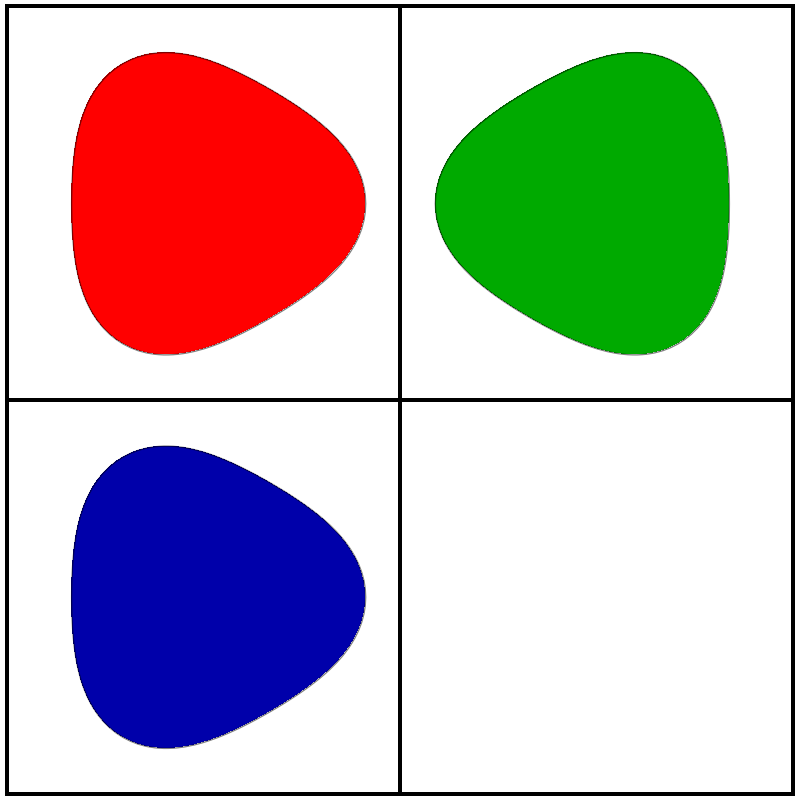}} &
\parbox[c][1.4cm][c]{1.2cm}{\centering \includegraphics[width=0.4cm]{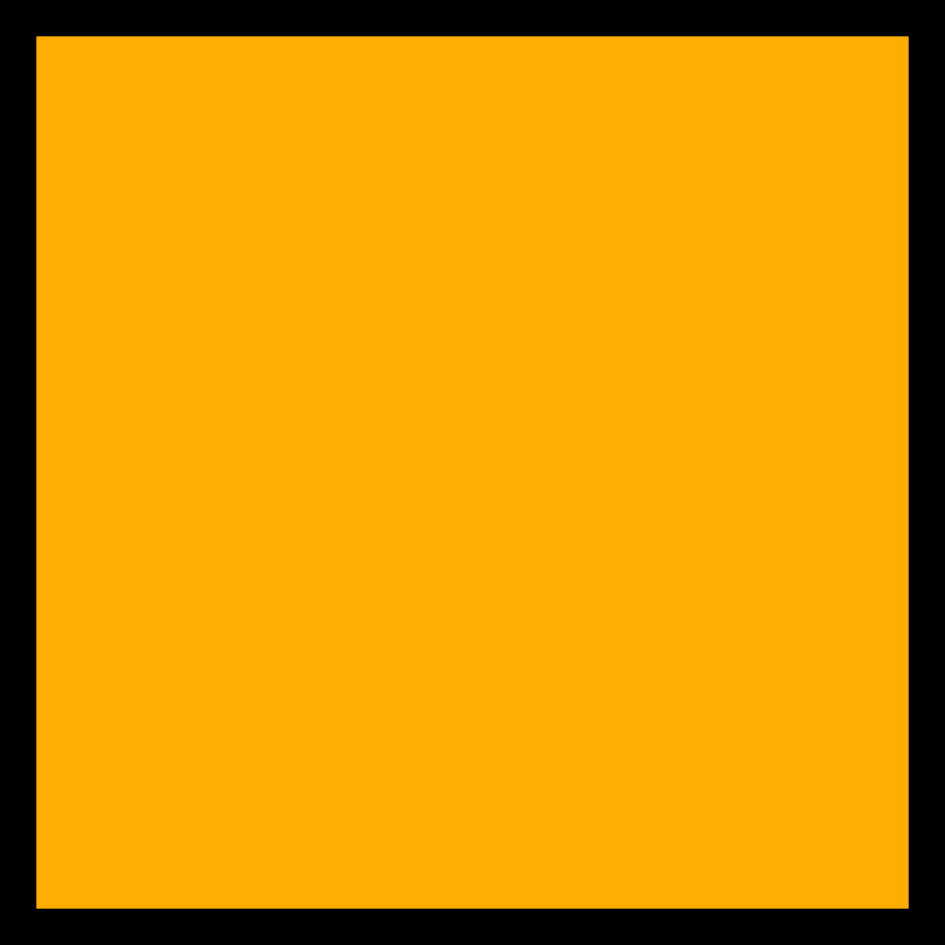}} \\
\hline \hline
\multicolumn{5}{|c|}{\parbox[c][0.6cm][c]{7.5cm}{\centering \textbf{Partially isospin polarized states}}} \\
\hline \hline

\parbox[c][1.1cm][c]{1.7cm}{\centering QM to HM} &
\parbox[c][1.1cm][c]{1cm}{\centering PIP$_{11}$} &
\parbox[c][1.1cm][c]{2cm}{\centering $\left(\alpha, \beta, 0, 0\right)$} &
\parbox[c][1.1cm][c]{1.5cm}{\centering \includegraphics[width=1cm]{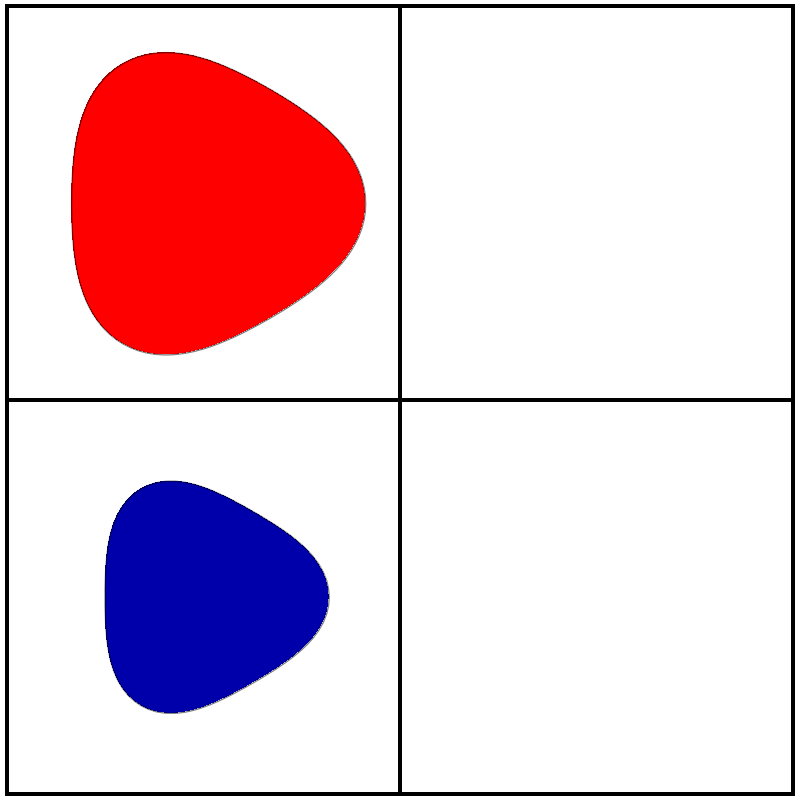}} &
\parbox[c][1.1cm][c]{1.2cm}{\centering \includegraphics[width=0.4cm]{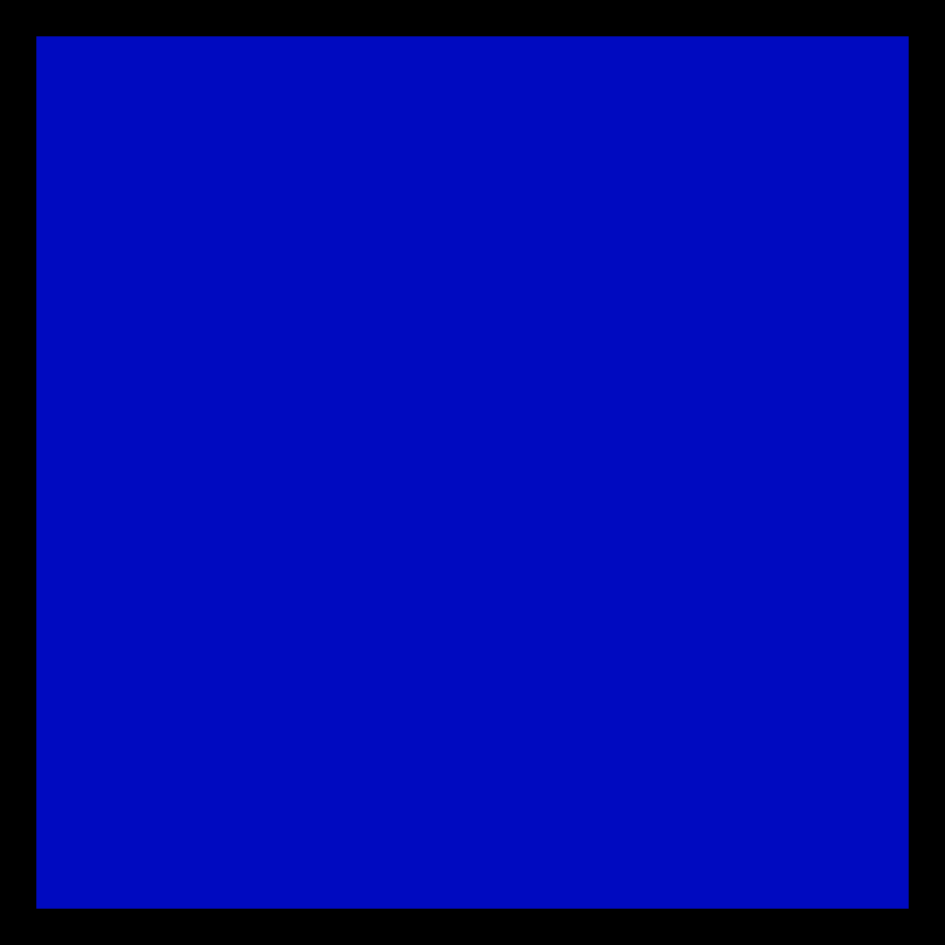}} \\
\hline

\parbox[c][1.1cm][c]{1.7cm}{\centering HM to FS} &
\parbox[c][1.1cm][c]{1cm}{\centering PIP$_{22}$} &
\parbox[c][1.1cm][c]{2cm}{\centering $\left(\alpha, \alpha, \beta, \beta\right)$} &
\parbox[c][1.1cm][c]{1.5cm}{\centering \includegraphics[width=1cm]{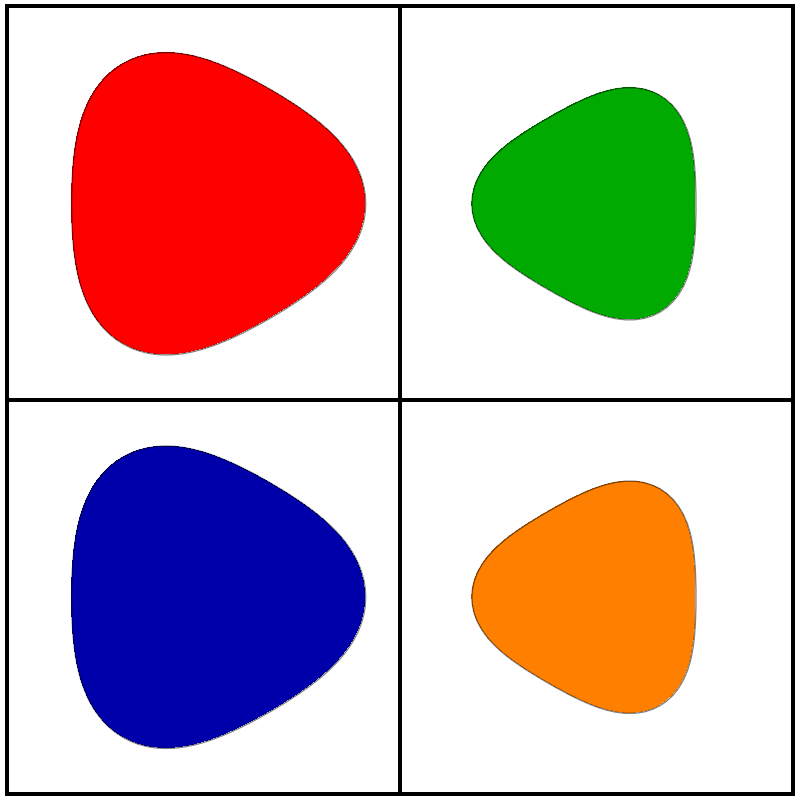}} &
\parbox[c][1.1cm][c]{1.2cm}{\centering \includegraphics[width=0.4cm]{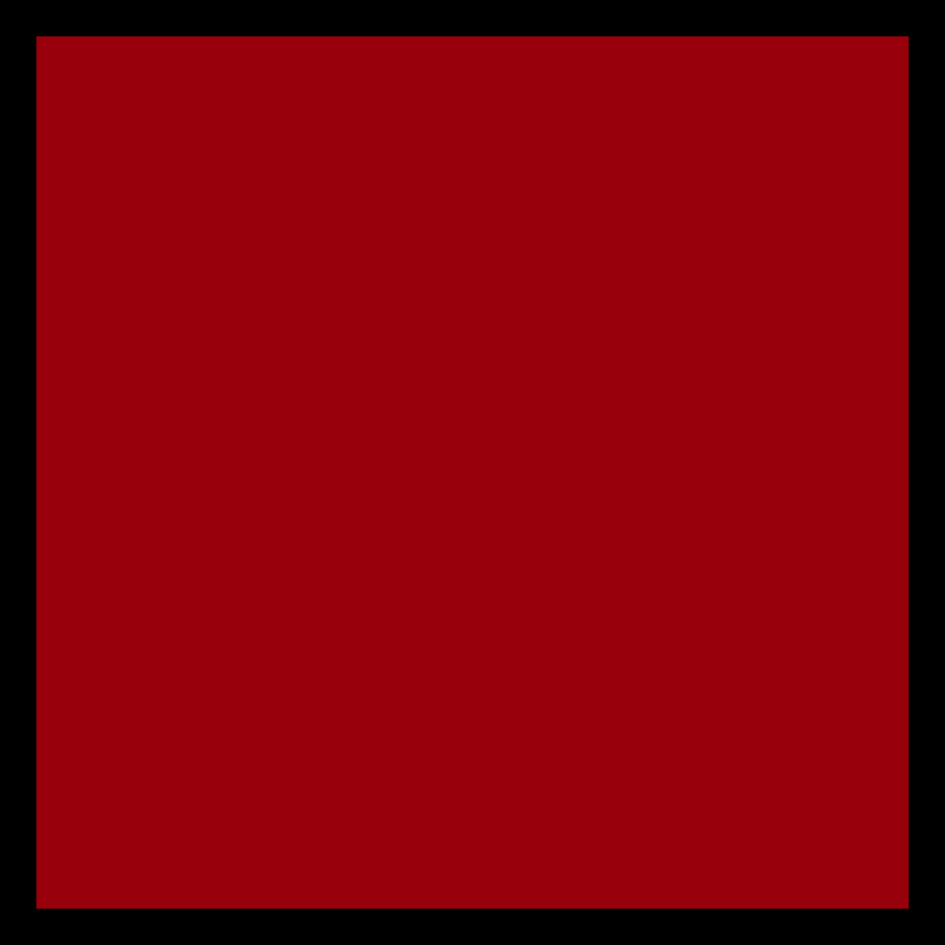}}\\
\hline

\parbox[c][1.1cm][c]{1.7cm}{\centering HM to TQM} &
\parbox[c][1.1cm][c]{1cm}{\centering PIP$_{21}$} &
\parbox[c][1.1cm][c]{2cm}{\centering $\left(\alpha, \alpha, \beta, 0\right)$} &
\parbox[c][1.1cm][c]{1.5cm}{\centering \includegraphics[width=1cm]{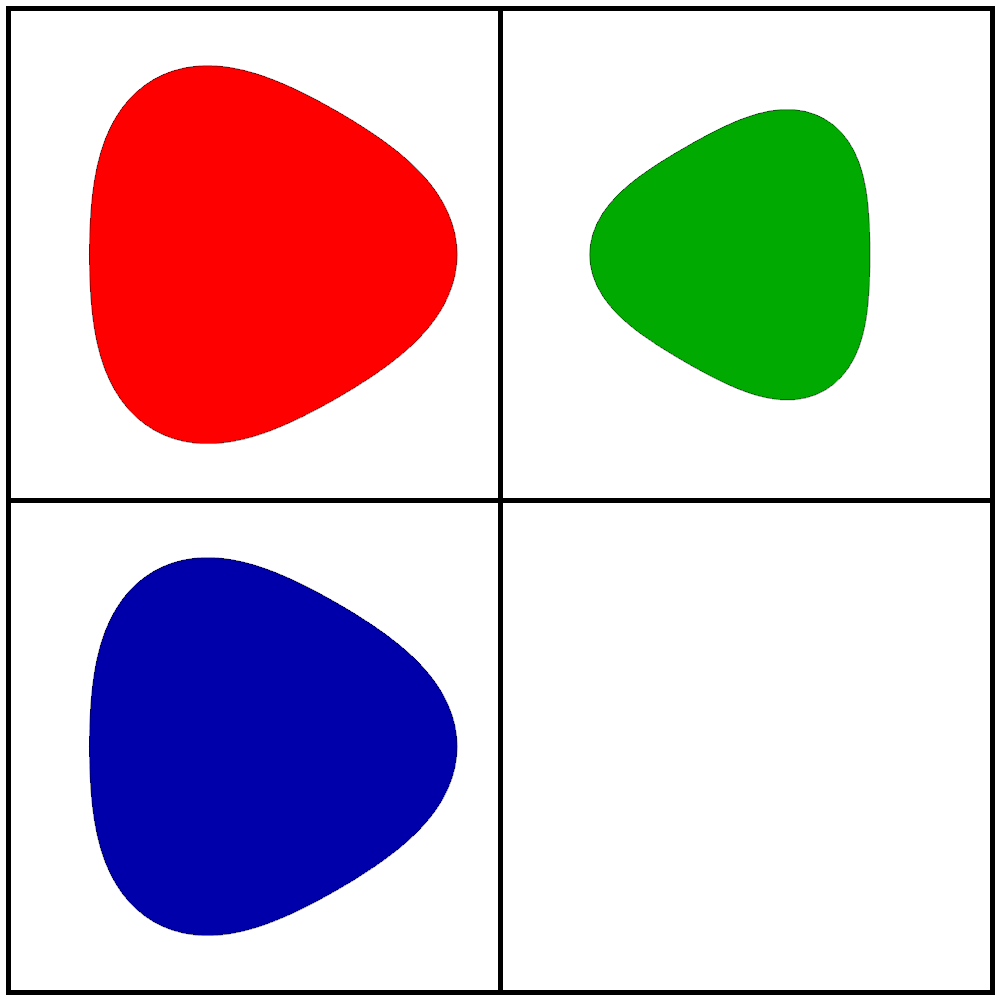}} &
\parbox[c][1.1cm][c]{1.2cm}{\centering \includegraphics[width=0.4cm]{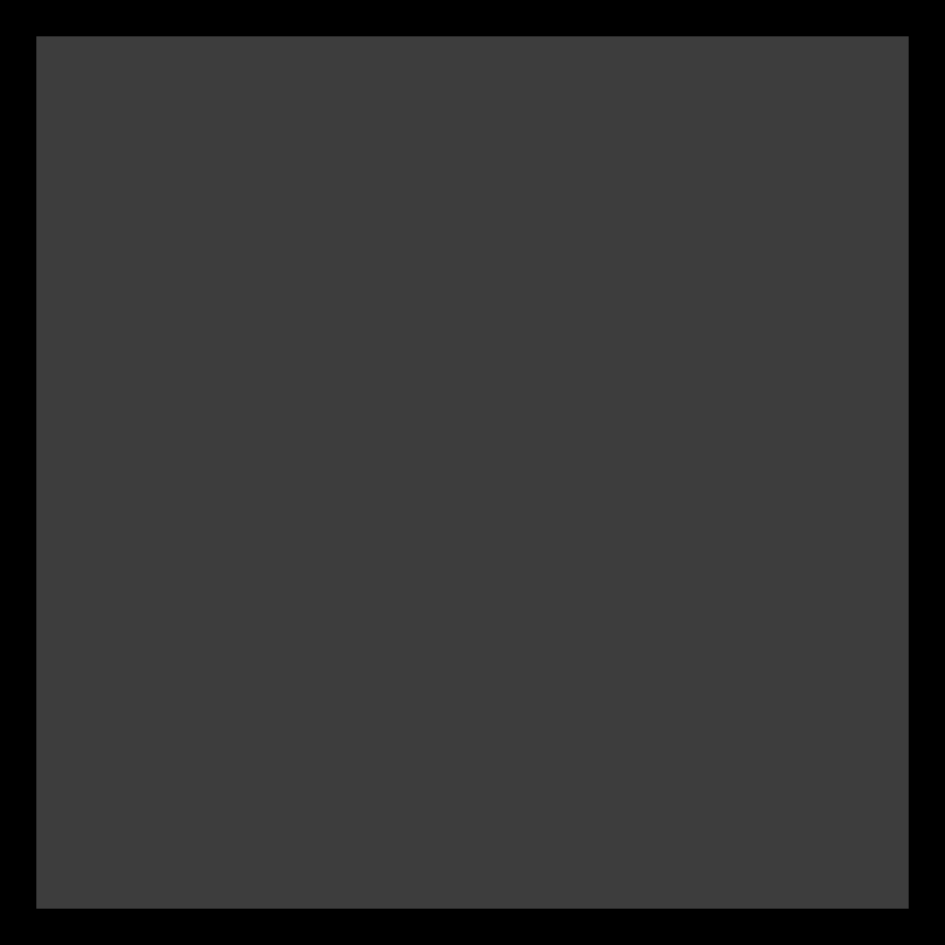}} \\
\hline

\parbox[c][1.1cm][c]{1.7cm}{\centering QM to TQM} &
\parbox[c][1.1cm][c]{1cm}{\centering PIP$_{12}$} &
\parbox[c][1.1cm][c]{2cm}{\centering $\left(\alpha, \beta, \beta, 0\right)$} &
\parbox[c][1.1cm][c]{1.5cm}{\centering \includegraphics[width=1cm]{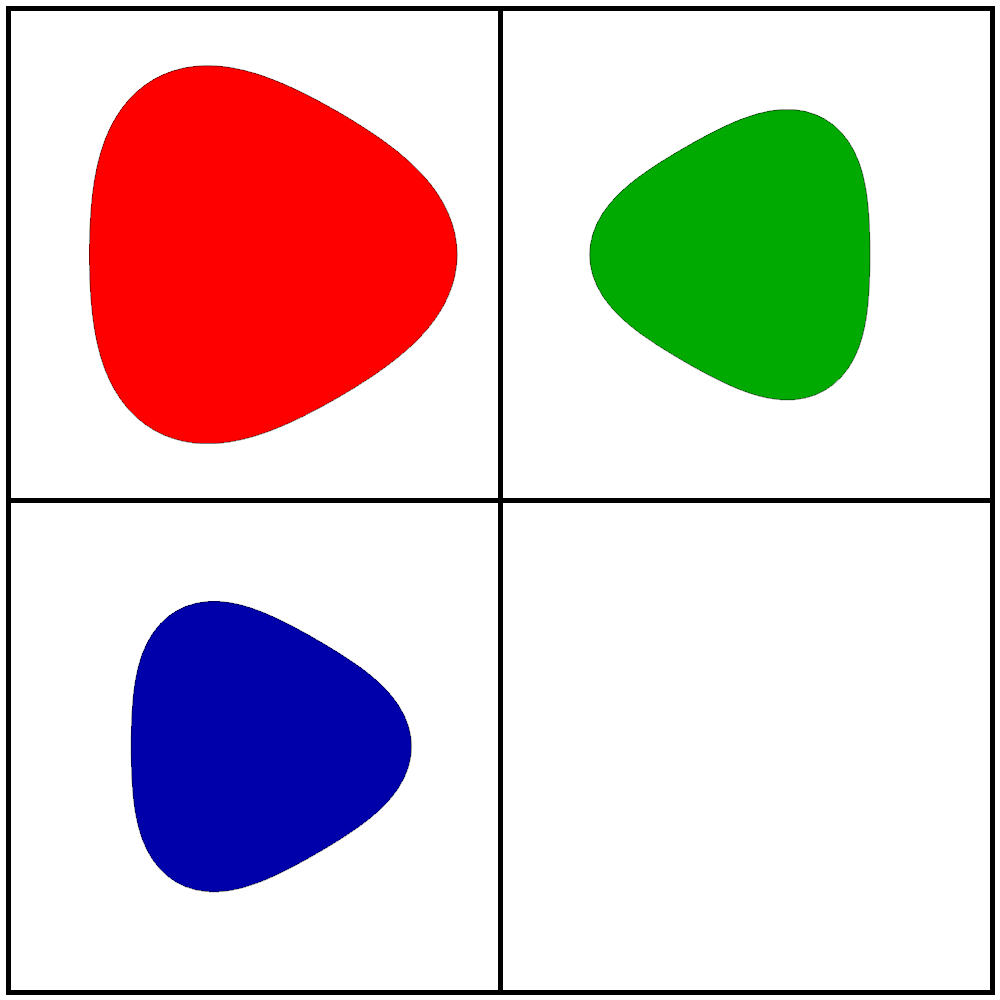}} &
\parbox[c][1.1cm][c]{1.2cm}{\centering \includegraphics[width=0.4cm]{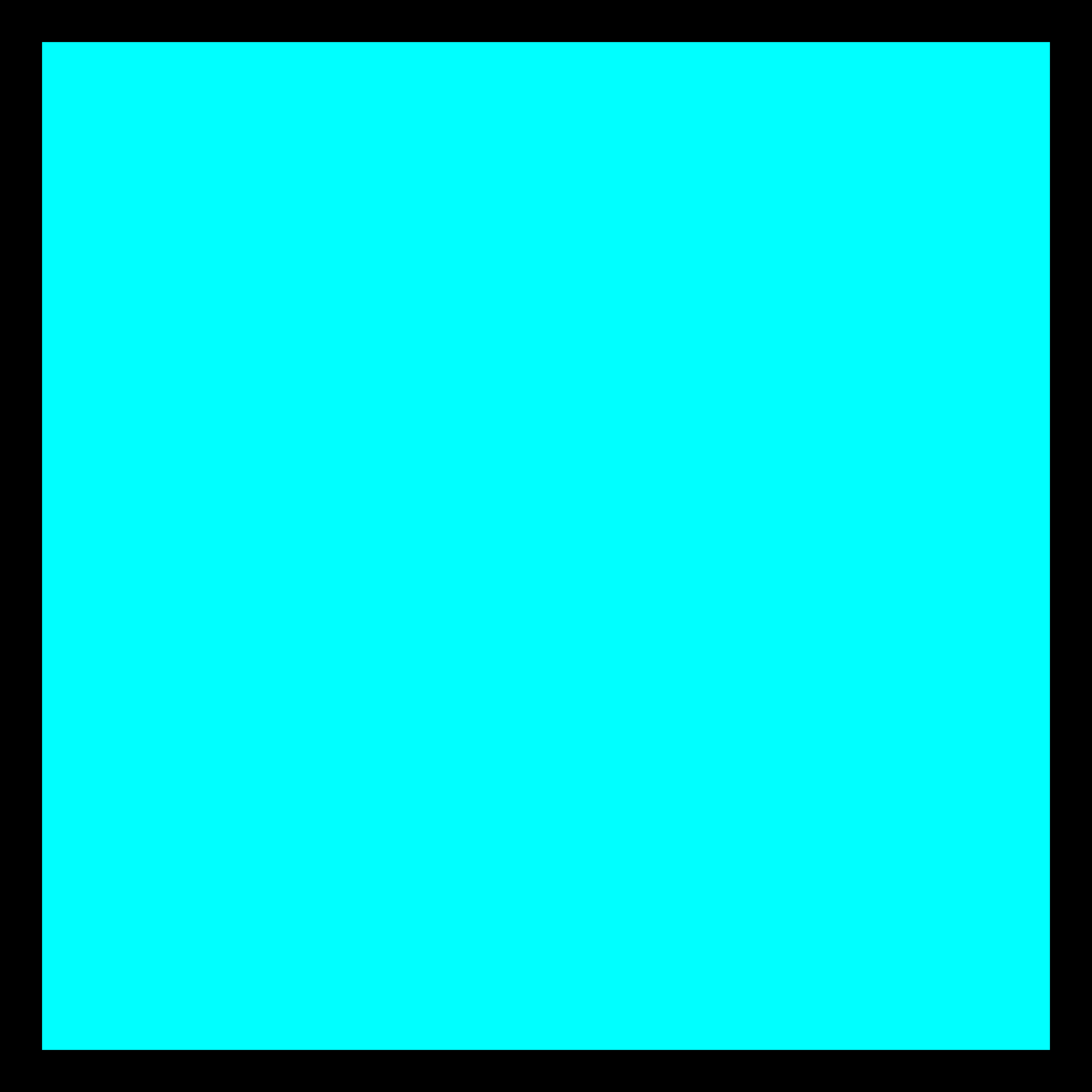}} \\
\hline

\parbox[c][1.1cm][c]{1.7cm}{\centering TQM to FS} &
\parbox[c][1.1cm][c]{1cm}{\centering PIP$_{31}$} &
\parbox[c][1.1cm][c]{2cm}{\centering $\left(\alpha, \alpha, \alpha, \beta\right)$} &
\parbox[c][1.1cm][c]{1.5cm}{\centering \includegraphics[width=1cm]{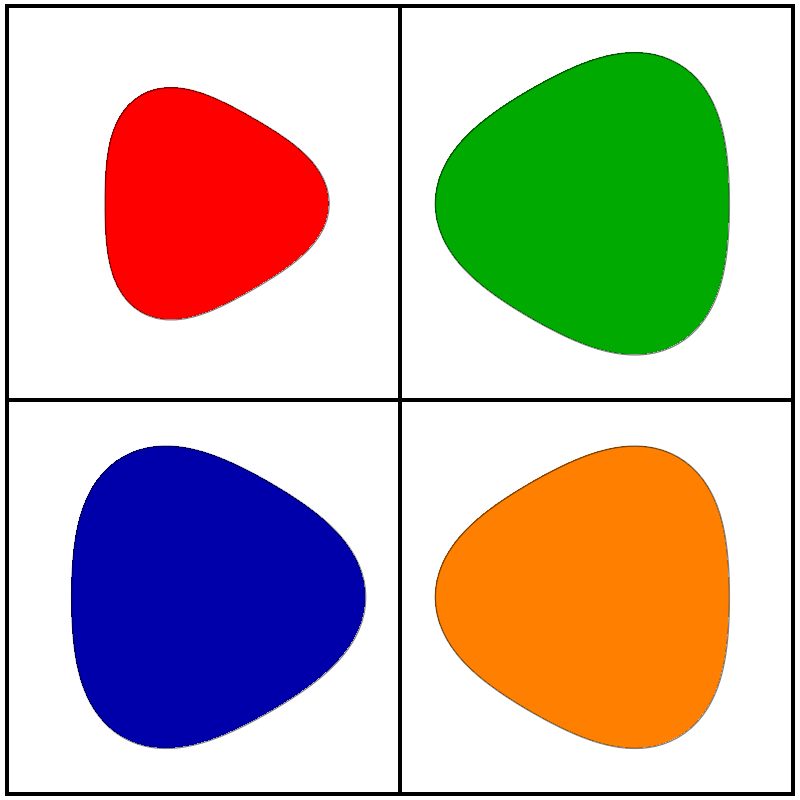}} &
\parbox[c][1.1cm][c]{1.2cm}{\centering \includegraphics[width=0.4cm]{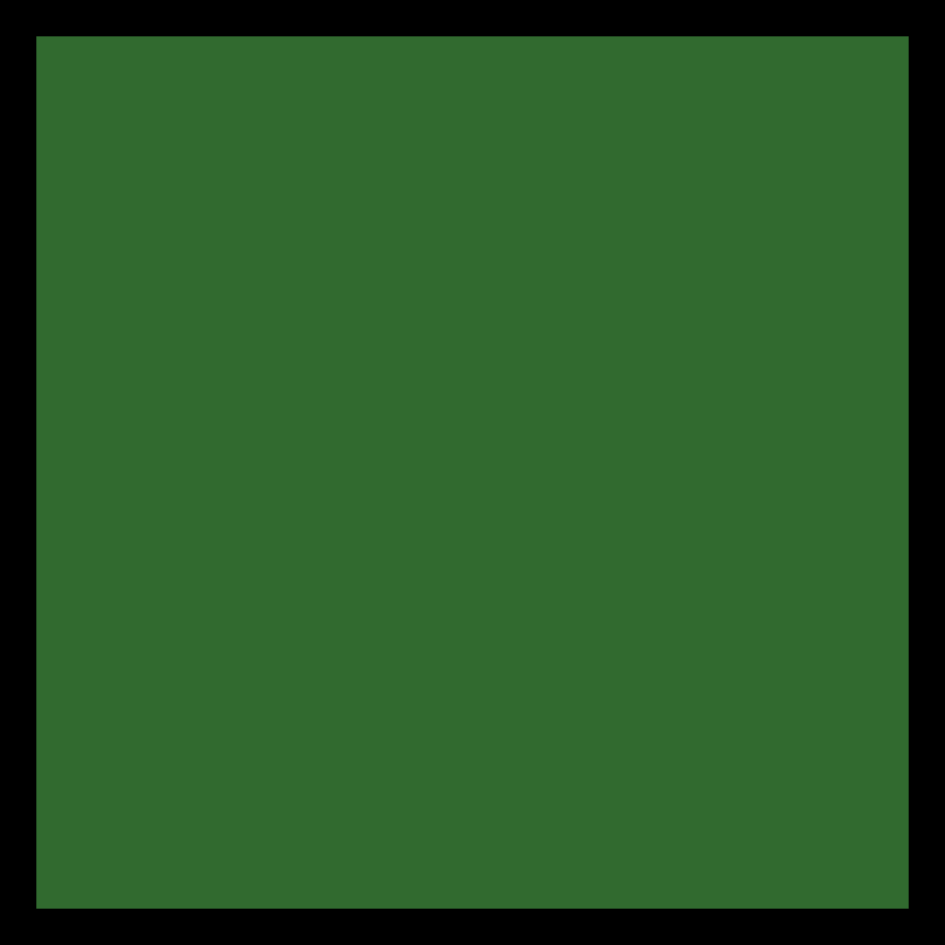}} \\
\hline

\parbox[c][1.1cm][c]{1.7cm}{\centering QM to FS} &
\parbox[c][1.1cm][c]{1cm}{\centering PIP$_{13}$} &
\parbox[c][1.1cm][c]{2cm}{\centering $\left(\alpha, \beta, \beta, \beta\right)$} &
\parbox[c][1.1cm][c]{1.5cm}{\centering \includegraphics[width=1cm]{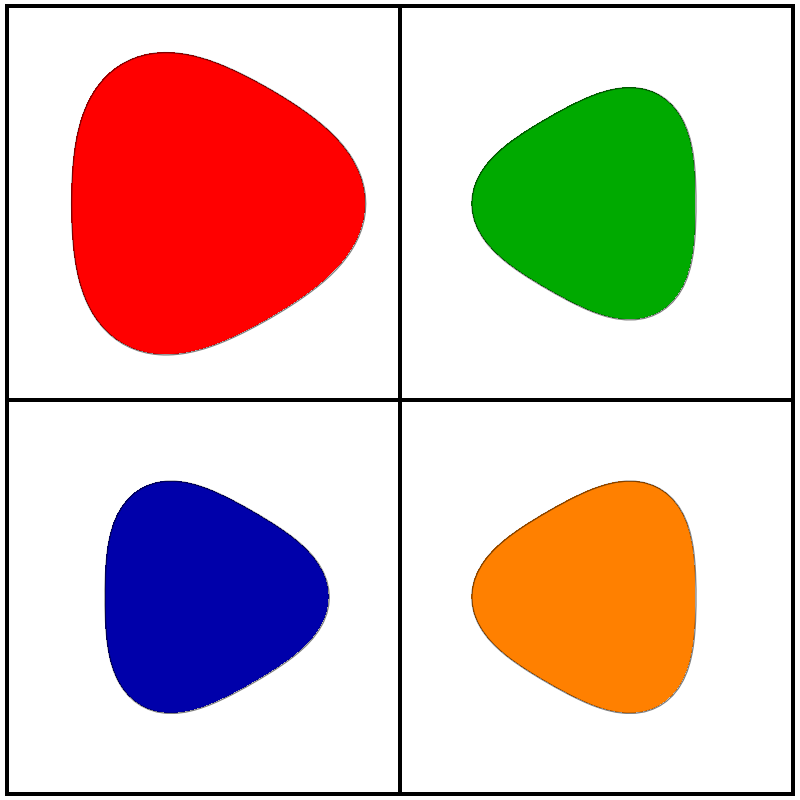}} &
\parbox[c][1.1cm][c]{1.2cm}{\centering \includegraphics[width=0.4cm]{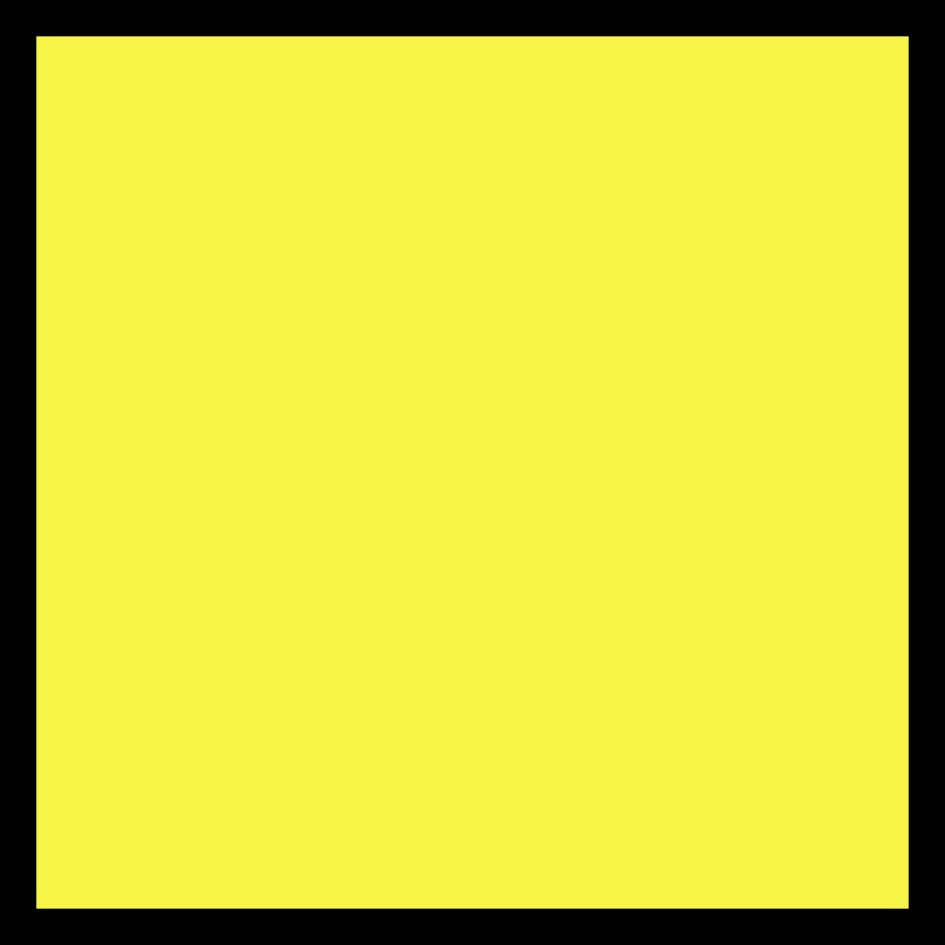}} \\
\hline

\end{tabular}
\caption{Classification of isospin-symmetry-broken phases. 
Top table: Sketch of the Isospin Polarized states where filled flavours are equally populated. 
Bottom table: Sketch of the Partially Isospin Polarized states where some flavours are  filled with a fraction of the most populated flavour. 
For all states are shown a representative Polarization vector and its Fermi surface.}
\label{tab:brokenSym_states}
\end{table}

\clearpage

\clearpage

\begin{table}
\centering
\begin{tabular}{cccccccccccc}
\hline
\hline
$\gamma_0$ && $\gamma_1$ && $\gamma_2$ && $\gamma_3$ &&  $\gamma_4$  && $\delta$  \\
\hline
3.1 && 0.38 && -0.015 && 0.29 && 0.141 && -0.0105  \\
\hline
\hline
\end{tabular}
\caption{Values for the hoppings of the continuum model in eV.}
\label{tab:hopping_params}
\end{table}

\end{document}